\documentclass[11pt]{article} \newcommand{\be}{\begin{equation}}
\newcommand{\ee}{\end{equation}}
\newcommand{\bea}{\begin{eqnarray}}
\newcommand{\eea}{\end{eqnarray}}
\newcommand{\sn}{{\rm sn}}

\newcommand{\dn}{{\rm dn}}
\newcommand{\cn}{{\rm cn}}
\newcommand{\sech}{{\rm sech}}

\begin{document}
\vspace{.5in} 
\begin{center} 
{\LARGE{\bf Higher Order Periodic Solutions of Coupled $\phi^4$ Models}} 
\end{center} 

\vspace{.3in}
\begin{center} 
{\LARGE{\bf Avinash Khare}} \\ 
{Institute of Physics, Bhubaneswar, Orissa 751005, India}
\end{center} 

\begin{center} 
{\LARGE{\bf Avadh Saxena}} \\ 
{Theoretical Division and Center for Nonlinear Studies, Los
Alamos National Laboratory, Los Alamos, NM 87545, USA}
\end{center} 

\vspace{.9in}
\noindent{\bf {Abstract:}}  

We obtain several higher order exact periodic  solutions of (i) a coupled 
symmetric $\phi^4$ model in an external field, (ii) an asymmetric coupled 
$\phi^4$ model, (iii) an asymmetric-symmetric coupled $\phi^4$ model, 
in terms of Lam\'e polynomials of order two  and obtain 
the corresponding hyperbolic solutions in the appropriate limit.  These 
solutions are unusual in the sense that while they are the solutions of 
the coupled problems, they are not the solutions of the 
corresponding uncoupled problems.

\newpage 
  
\section{Introduction} 
Coupled double well ($\phi^4$) one-dimensional potentials are prevalent 
in both condensed matter physics and field theory.  Few examples of 
current interest include spin configurations, domain walls and magnetic 
phase transitions in multiferroic materials \cite{flp,cm} and $\omega$ 
phase transition in various elements and alloys \cite{ss}. In two recent 
publications \cite{ks1,ks2}, we obtained a large number of periodic 
solutions, in terms of the Lam\'e polynomials of order one \cite{gr,finkel}, 
for (i) a coupled symmetric $\phi^4$ model in an external field and (ii) 
an asymmetric coupled $\phi^4$ model, both models with a biquadratic 
coupling. All those solutions had the feature that in the uncoupled limit, 
they reduce to the well known solutions of the uncoupled symmetric or 
asymmetric double well problem as the case may be. The purpose of this 
paper is to point out that both these coupled models have, in addition, 
truly novel solutions, in terms of the Lam\'e polynomials of order two, which 
only exist due to the presence of the coupling between the two fields. In 
other words, while the Lam\'e polynomials of order two are the solutions of 
the coupled problem, they are {\it not} the solutions of the decoupled 
problem. For completeness, we also consider an asymmetric-symmetric coupled
$\phi^4$ model (which corresponds to a first order transition in one field 
and a second order transition in the other) and show that not only Lam\'e 
polynomials of order one but also Lam\'e polynomials of order two are the 
solutions of this coupled problem, even though only Lam\'e polynomials of 
order one are the solutions of the uncoupled problem.  This model is 
relevant for certain martensitic transformations in elements 
\cite{rochal,toledano}   
  
The paper is organized as follows. In Sec. II we provide the novel 
periodic as well as the corresponding hyperbolic solutions for the 
coupled symmetric $\phi^4$ model \cite{ks1} with an explicit biquadratic 
coupling in the presence of an external field (with an additional 
linear-quadratic coupling) \cite{ks1}. Note that the symmetric $\phi^4$ 
model, in the decoupled limit, corresponds to a second order transition 
in both the fields. We show that while the solutions of the uncoupled 
$\phi^4$ problem are the Lam\'e polynomials of order one, (i.e. $\sn$, $\cn$, 
$\dn$), for the coupled problem, not only the Lam\'e polynomials of order 
one \cite{ks1}, but even the Lam\'e polynomials of order two are the 
solutions of the 
coupled field equations. In Sec. III we provide the periodic (Lam\'e 
polynomials of order two) and the corresponding hyperbolic solutions for 
the coupled asymmetric $\phi^4$ model, which corresponds to a first order 
transition in both the fields \cite{ks2}.  In Sec. IV and V we consider the
Lam\'e polynomial solutions of order one and two respectively of a coupled
asymmetric-symmetric $\phi^4$ model. Finally, we conclude in Sec. 
VI with summary and possible extensions.

\section{Coupled symmetric $\phi^4$ model in an external field}

In \cite{ks1} we had considered the following potential, with a 
{\it biquadratic} coupling between the two fields 
and in an external magnetic field ($H_z$)  
\be\label{1}
V=\alpha_1 \phi^2 +\beta_1 \phi^4 +\alpha_2 \psi^2 +\beta_2 \psi^4 
+\gamma\phi^2 \psi^2
-H_z[\rho_1 \phi+\rho_2 \phi^3+\rho_3 \phi \psi^2]\,, 
\ee
where $\alpha_i$, $\beta_i$, $\gamma$ and $\rho_i$ are material (or 
system) dependent parameters.  For $\alpha_1<0,\alpha_2 <0$ and
$\beta_1>0,\beta_2>0$, this model corresponds to second order transitions
in both fields $\phi$ and $\psi$. The corresponding (static) equations 
of motion are  
\be\label{2}
\frac{d^2 \phi}{dx^2}=2\alpha_1 \phi +4\beta_1 \phi^3 
+2\gamma\phi \psi^2
-H_z[\rho_1+3\rho_2 \phi^2+\rho_3 \psi^2]\,,
\ee
\be\label{3}
\frac{d^2 \psi}{dx^2}=2\alpha_2 \psi +4\beta_2 \psi^3 
+2\gamma\phi^{2} \psi
-2H_z\rho_3 \phi \psi\,.
\ee
These coupled set of equations admit several novel periodic solutions 
(i.e. Lam\'e polynomials of order 2), which 
we now discuss one by one systematically. 

\subsection{Solution I}

It is not difficult to show that
\be\label{1r}
\phi =F+A\sn^2[D(x+x_0),m]\,,~~\psi =G+B\sn^2[D(x+x_0),m]\,,
\ee
is an exact solution to coupled field Eqs. (\ref{2}) and (\ref{3}) provided 
the following eight field equations are satisfied
\be\label{1r1}
2\alpha_1 F+4\beta_1 F^3+2\gamma FG^2-H_z\rho_1-3H_z \rho_2F^2
-H_z\rho_3G^2 =2AD^2\,,
\ee
\be\label{1r2}
2\alpha_1 A+12\beta_1 F^2 A+4\gamma BFG+2\gamma AG^2
-6H_z \rho_2 AF-2H_z\rho_3 BG =-4(1+m)AD^2\,,
\ee
\be\label{1r3}
12\beta_1 F A^2+2\gamma FB^2+4\gamma ABG-3H_z \rho_2A^2
-H_z\rho_3B^2 =6AmD^2\,,
\ee
\be\label{1r4}
2\beta_1 A^2+\gamma B^2=0\,,
\ee
\be\label{1r5} 
2\alpha_2 G+4\beta_2 G^3+2\gamma GF^2-2H_z\rho_3GF =2BD^2\,,
\ee
\be\label{1r6}
2\alpha_2 B+12\beta_2 G^2 B+4\gamma AFG+2\gamma BF^2
-2H_z\rho_3 (BF+AG) =-4(1+m)BD^2\,,
\ee
\be\label{1r7}
12\beta_2 G B^2+2\gamma GA^2+4\gamma ABF
-2H_z\rho_3AB =6mBD^2\,,
\ee
\be\label{1r8}
2\beta_2 B^2+\gamma A^2=0\,.
\ee
Here $A$ and $B$ denote the amplitudes of the ``pulse lattice", $F$ 
and $G$ are constants, $D$ is an inverse characteristic length and 
$x_0$ is the (arbitrary) location of the pulse.  Five of these equations 
determine the five unknowns $A,B,D,F,G$ while the other three equations, 
give three constraints between the nine parameters $\alpha_{1,2},\beta_{1,2}, 
\gamma,H_z,\rho_1, \rho_2,\rho_3$. 
In particular, from Eqs. (\ref{1r4}) and (\ref{1r8}) it follows that 
\be\label{1.1}
\gamma<0\,,~~|\gamma|^2=4\beta_1 \beta_2\,, ~~~\sqrt{\beta_1}A^2
=\sqrt{\beta_2}B^2\,.
\ee
Few comments are in order at this stage.

\begin{enumerate}
 
\item From the Eq. (\ref{1r5}) it follows that no solution of form (\ref{1r}) 
exists in case $G=0$. Thus no solutions exist with $\psi=B\sn^2[D(x+x_0),m]$ 
irrespective of the value of $F$.  In fact one can also show that no solution 
exists in case $B=-G$ or if $B=-mG$ unless $m=1$. In other words, even the 
solutions of the form $\psi=G \cn^2[D(x+x_0),m]$ or $\psi=G\dn^2[D(x+x_0),m]$ 
do not exist, no matter what $F$ is, except when $m=1$. 

\item In the special case of $H_z=0$, the field equations (\ref{1r1}) to
(\ref{1r8}) are completely symmetrical in $\phi$ and $\psi$. It is easily 
shown that in this case, solution (\ref{1r}) does not exist. 

\end{enumerate}

{\bf Solution at $m=1$}: In the special case of $m=1$, the solution (\ref{1r}) 
goes over to the hyperbolic nontopological soliton solution
\be\label{1s}
\phi =F+A\tanh^2[D(x+x_0)]\,,~~\psi =G+B\tanh^2[D(x+x_0)]\,,
\ee
provided the field Eqs. (\ref{1r1}) to (\ref{1r8}) with $m=1$ are satisfied.
This hyperbolic soliton solution takes particularly simple form in two cases
which we mention one by one. 

(i) {\bf $F=0,\, G=-B$}: In this case, the nontopological soliton 
solution (\ref{1s}) takes the simpler form
\be\label{1s1}
\phi =A\tanh^2[D(x+x_0)]\,,~~\psi =B\sech^2[D(x+x_0)]\,.
\ee
By analyzing Eqs. (\ref{1r1}) to (\ref{1r8}) it is easily shown that such a
solution exists provided $\gamma <0\,,\alpha_2<0\,,\rho_3<0$. Further, while
Eqs. (\ref{1r4}) and (\ref{1r8}) still continue to hold good, the other field
equations take slightly simpler form
\be\label{1s2}
D^2=|\alpha_2|-|\gamma|A^2\,,
\ee
\be\label{1s3}
3\alpha_2=H_z|\rho_3|A+|\gamma|A^2\,,
\ee
\be\label{1s4}
2AD^2==H_z\rho_1+H_z|\rho_3|B^2\,,
\ee
\be\label{1s5}
H_z\rho_1+3H_z\rho_2A^2=2\alpha_1A+4\beta_1A^3\,,
\ee
\be\label{1s6}
3\alpha_1A+10\beta_1A^3=6H_Z\rho_2A^2+H_z|\rho_3|B^2\,.
\ee

(ii) {\bf $F=-A,\, G=-B$}: In this limit the nontopological soliton 
solution (\ref{1s}) takes the simpler form
\be\label{1t1}
\phi =A\sech^2[D(x+x_0)]\,,~~\psi =B\sech^2[D(x+x_0)]\,.
\ee
provided field Eqs. (\ref{1r1}) and (\ref{1r5}) hold good
and further 
\be\label{1t2}
\rho_1=0\,,~~\gamma <0\,,~~\alpha_1=\alpha_2>0\,,<0\,,~~\rho_2>0\,,~
\rho_3>0\,,
\ee
\be\label{1t3}
D^2=\frac{\alpha_1}{2}\,,~~A=\frac{3\alpha_1}{2H_z\rho_3}\,,~~
3\rho_2A^2=(2A^2-B^2)\rho_3\,.
\ee

\subsection{Solution II}

It is not difficult to show that
\be\label{2r}
\phi =F+A\sn^2[D(x+x_0),m]\,,~~\psi =B\sn[D(x+x_0),m]\cn[D(x+x_0),m]\,,
\ee
is an exact solution to coupled field Eqs. (\ref{2}) and (\ref{3}) provided 
the following seven field equations are satisfied
\be\label{2r1}
2\alpha_1 F+4\beta_1 F^3-H_z\rho_1-3H_z \rho_2F^2 =2AD^2\,,
\ee
\be\label{2r2}
2\alpha_1 A+12\beta_1 F^2 A+2\gamma FB^2
-6H_z \rho_2 AF-H_z\rho_3 B^2 =-4(1+m)AD^2\,,
\ee
\be\label{2r3}
12\beta_1 F A^2+2\gamma B^2 (A-F)-3H_z \rho_2A^2
+H_z\rho_3B^2 =6AmD^2\,,
\ee
\be\label{2r4}
2\beta_1 A^2-\gamma B^2=0\,,
\ee
\be\label{2r5} 
2\alpha_2 +2\gamma F^2-2H_z\rho_3F =-(4+m)D^2\,,
\ee
\be\label{2r6}
4\beta_2 B^2+4\gamma AF
-2H_z\rho_3A =6mD^2\,,
\ee
\be\label{2r7}
2\beta_2 B^2-\gamma A^2=0\,.
\ee
Four of these equations determine the four unknowns $A,B,D,F$ while the 
other three equations, give three constraints between the nine parameters 
$\alpha_{1,2},\beta_{1,2}, \gamma,H_z,\rho_1, \rho_2,\rho_3$. 
In particular, from Eqs. (\ref{2r4}) and (\ref{2r7}) it follows that 
\be\label{2.1}
\gamma>0\,,~~\gamma^2=4\beta_1 \beta_2\,, ~~~\sqrt{\beta_1}A^2
=\sqrt{\beta_2}B^2\,.
\ee
It is easy to show that the solution (\ref{2r}) continues to exist at  
$F=0,-A,-A/m$ so long as $H_z \ne 0$.

{\bf Solution at $m=1$}: In the special case of $m=1$, the solution (\ref{2r}) 
goes over to the hyperbolic nontopological soliton solution
\be\label{2.2}
\phi =F+A\tanh^2[D(x+x_0)]\,,~~\psi =B\tanh[D(x+x_0)]\sech[D(x+x_0)]\,,
\ee 
provided the field Eqs. (\ref{2r1}) to (\ref{2r7}) with $m=1$ are satisfied.

{\bf Special case $H_z=0$}: In the special case of $H_z=0$, the field equations
(\ref{2}) and (\ref{3}) are symmetrical in $\phi$ and $\psi$. In this
case, the Eqs. (\ref{2r1}) to (\ref{2r7}) take rather simple form. In 
particular, in case $H_z=0$, it is easily shown that the solution (\ref{2r}) 
exists provided
\be\label{2s1}
\gamma=2\beta_1=2\beta_2\,,~~A^2=B^2\,,
\ee
while the remaining equations take the simpler form
\be\label{2s2}
3mD^2=(1+2x)\gamma A^2\,,
\ee
\be\label{2s3}
D^2=\alpha_1 x+\gamma A^2 x^3\,,
\ee
\be\label{2s4}
-2(1+m)D^2=\alpha_1 +x(1+3x)\gamma A^2\,,
\ee
\be\label{2s5}
-(4+m)D^2=2\alpha_2+2x^2 \gamma A^2\,,
\ee
where $x=F/A$. On solving these equations, one finds that the only acceptable
solution is given by
\be\label{2s6}
3mx=-(1+m)+\sqrt{1-m+m^2}\,,
\ee
using which one can then easily express $D^2, \alpha_2$ and $\gamma A^2$ in 
terms of $\alpha_1$. 

In particular, at $H_z=0$ and $m=1$, the solution (\ref{2.2}) exists provided
relations (\ref{2s1}) are satisfied and further
\be\label{2s7}
F=-\frac{A}{3}\,,~~\alpha_1<0\,,~~\alpha_2<0\,,~~D^2=\frac{|\alpha_1|}{4}\,,
~~\gamma A^2 =\frac{3|\alpha_1|}{2}\,,~~|\alpha_2|=\frac{7}{8}\alpha_1\,.
\ee

\subsection{Solution III}

It is not difficult to show that unlike the solution (\ref{2r}), the
solution
\be\label{3r}
\phi =B\sn[D(x+x_0),m]\cn[D(x+x_0),m]\,,~~
\psi =F+A\sn^2[D(x+x_0),m]\,,
\ee
exists only if $H_z=0$. But since at $H_z=0$, the field Eqs. (\ref{2}) and 
(\ref{3}) are symmetrical in $\phi$ and $\psi$, hence (\ref{3r})
is a solution to field Eqs. (\ref{2}) and (\ref{3}) provided 
Eqs. (\ref{2s1}) to (\ref{2s7}) (with suitable change of parameters) are 
satisfied.

\subsection{Solution IV}

It is not difficult to show that
\be\label{4r}
\phi =F+A\sn^2[D(x+x_0),m]\,,~~\psi =B\sn[D(x+x_0),m]\dn[D(x+x_0),m]\,,
\ee
is an exact solution to coupled field Eqs. (\ref{2}) and (\ref{3}) provided 
the following seven field equations are satisfied
\be\label{4r1}
2\alpha_1 F+4\beta_1 F^3-H_z\rho_1-3H_z \rho_2F^2 =2AD^2\,,
\ee
\be\label{4r2}
2\alpha_1 A+12\beta_1 F^2 A+2\gamma FB^2
-6H_z \rho_2 AF-H_z\rho_3 B^2 =-4(1+m)AD^2\,,
\ee
\be\label{4r3}
12\beta_1 F A^2+2\gamma B^2 (A-Fm)-3H_z \rho_2A^2
+mH_z\rho_3B^2 =6AmD^2\,,
\ee
\be\label{4r4}
2\beta_1 A^2-m\gamma B^2=0\,,
\ee
\be\label{4r5} 
2\alpha_2 +2\gamma F^2-2H_z\rho_3F =-(1+4m)D^2\,,
\ee
\be\label{4r6}
4\beta_2 B^2+4\gamma AF
-2H_z\rho_3A =6mD^2\,,
\ee
\be\label{4r7}
2m\beta_2 B^2-\gamma A^2=0\,.
\ee
Four of these equations 
determine the four unknowns $A,B,D,F$ while the other three equations, 
give three constraints between the nine parameters $\alpha_{1,2},\beta_{1,2}, 
\gamma,H_z,\rho_1, \rho_2,\rho_3$. 
In particular, from Eqs. (\ref{4r4}) and (\ref{4r7}) it follows that 
\be\label{4.1}
\gamma>0\,,~~\gamma^2=4\beta_1 \beta_2\,, ~~~\sqrt{\beta_1}A^2
=m\sqrt{\beta_2}B^2\,.
\ee
It is easy to show that the solution (\ref{4r}) continues to exist in case  
$F=0,-A,-A/m$ as long as $H_z \ne 0$.

{\bf Solution at $m=1$}: In the special case of $m=1$, the solution (\ref{4r}) 
goes over to the hyperbolic nontopological soliton solution (\ref{2.2})
provided the field Eqs. (\ref{4r1}) to (\ref{4r7}) with $m=1$ are satisfied.

{\bf Special case $H_z=0$}: In the special case of $H_z=0$, the field equations
(\ref{2}) and (\ref{3}) are symmetrical in $\phi$ and $\psi$. In this
case, Eqs. (\ref{4r1}) to (\ref{4r7}) take rather simple form. 
In particular, in case $H_z=0$, it is easily shown that the 
solution (\ref{4r}) exists provided Eq. (\ref{2s1}) is satisfied
while the remaining equations take the simpler form
\be\label{4s2}
3m^2D^2=(1+2mx)\gamma A^2\,,
\ee
\be\label{4s3}
D^2=\alpha_1 x+\gamma A^2 x^3\,,
\ee
\be\label{4s4}
-2(1+m)mD^2=m\alpha_1 +x(1+3mx)\gamma A^2\,,
\ee
\be\label{4s5}
-(1+4m)D^2=2\alpha_2+2x^2 \gamma A^2\,,
\ee
where $x=F/A$. On solving these equations, one finds that the only acceptable
solution is the one with $x$ again given by Eq. (\ref{2s6})
using which one can then easily express $D^2, \alpha_2$ and $\gamma A^2$ in 
terms of $\alpha_1$. 
At $H_z=0$ and $m=1$, of course the solution goes over to the solution 
(\ref{2.2}), which exists provided
relations (\ref{2s1}) to (\ref{2s7}) are satisfied.

\subsection{Solution V}

It is not difficult to show that unlike the solution (\ref{4r}), the
solution
\be\label{5r}
\phi =B\sn[D(x+x_0),m]\dn[D(x+x_0),m]\,,~~
\psi =F+A\sn^2[D(x+x_0),m]\,,
\ee
exists only if $H_z=0$. But since at $H_z=0$, the field Eqs. (\ref{2}) and 
(\ref{3}) are symmetrical in $\phi$ and $\psi$, hence (\ref{5r})
is a solution to field Eqs. (\ref{2}) and (\ref{3}) provided 
Eqs. (\ref{4s2}) to (\ref{4s5}) (with suitable change of parameters) are 
satisfied.

\subsection{Solution VI}

It is not difficult to show that
\be\label{6r}
\phi =F+A\sn^2[D(x+x_0),m]\,,~~\psi =B\cn[D(x+x_0),m]\dn[D(x+x_0),m]\,,
\ee
is an exact solution to coupled field Eqs. (\ref{2}) and (\ref{3}) provided 
the following seven field equations are satisfied
\be\label{6r1}
2\alpha_1 F+4\beta_1 F^3+2\gamma B^2 F-H_z\rho_1-3H_z \rho_2F^2 
-H_z \rho_3 B^2 =2AD^2\,,
\ee
\be\label{6r2}
2\alpha_1 A+12\beta_1 F^2 A+2\gamma B^2[A-(1+m)F]
-6H_z \rho_2 AF+(1+m)H_z\rho_3 B^2 =-4(1+m)AD^2\,,
\ee
\be\label{6r3}
12\beta_1 F A^2+2\gamma B^2 [mF-(1+m)A]-3H_z \rho_2A^2
-mH_z\rho_3B^2 =6AmD^2\,,
\ee
\be\label{6r4}
2\beta_1 A^2+\gamma B^2=0\,,
\ee
\be\label{6r5} 
2\alpha_2 +4\beta_2 B^2+2\gamma F^2-2H_z\rho_3F =-(1+m)D^2\,,
\ee
\be\label{6r6}
-4(1+m)\beta_2 B^2+4\gamma AF
-2H_z\rho_3A =6mD^2\,,
\ee
\be\label{6r7}
2m\beta_2 B^2+\gamma A^2=0\,.
\ee
Four of these equations 
determine the four unknowns $A,B,D,F$ while the other three equations, 
give three constraints between the nine parameters $\alpha_{1,2},\beta_{1,2}, 
\gamma,H_z,\rho_1, \rho_2,\rho_3$. 
In particular, from above Eqs. (\ref{6r4}) and (\ref{6r7}) it follows that 
\be\label{6.1}
\gamma<0\,,~~|\gamma|^2=4m\beta_1 \beta_2\,, ~~~\sqrt{\beta_1}A^2
=\sqrt{m\beta_2}B^2\,.
\ee
It is easily shown that the solution (\ref{6r}) continues to exist in case 
$F=0,-A,-A/m$ as long as $H_z \ne 0$.

{\bf Solution at $m=1$}: In the special case of $m=1$, the solution (\ref{6r}) 
goes over to the hyperbolic nontopological soliton solution 
\be\label{6.2}
\phi =F+A\tanh^2[D(x+x_0)]\,,~~\psi =B\sech^2[D(x+x_0)]\,,
\ee 
which is essentially the solution (\ref{1s}) with $B=-G$
provided the field Eqs. (\ref{6r1}) to (\ref{6r7}) with $m=1$ are satisfied.

{\bf Special case $H_z=0$}: In the special case of $H_z=0$, the field equations
(\ref{2}) and (\ref{3}) are symmetrical in $\phi$ and $\psi$. In this
case, at least at $m=1$, one can show that there is no solution to these
equations. Of course this is expected since we know 
from the discussion of solution (\ref{1r}) that at $m=1$ and $H_z=0$, 
solution (\ref{6.2}) does not exist. 

\section{Coupled Asymmetric $\phi^4$ Model}

Recently we had also considered a coupled asymmetric $\phi^4$ model 
\cite{ks2}, which in the uncoupled limit corresponds to a first order 
transition in both the fields, and had obtained periodic solutions in 
terms of the Lam\'e polynomials of order one.  The purpose of this section 
is to show that the Lam\'e polynomials of order two also constitute exact 
solutions of the same model. It is worth pointing out here that while 
the Lam\'e polynomials of order one are also solutions of the asymmetric 
uncoupled model, the Lam\'e polynomials of order two are in fact {\it not} 
the solutions of the uncoupled equations, thereby giving us genuinely 
new solutions of the coupled problem.

The potential that we considered in \cite{ks2} is given by ($\beta_1 >0
,\beta_2 > 0$)
\be\label{3.1}
V=\alpha_1 \phi^2 +\delta_1 \phi^3+\beta_1 \phi^4 +\alpha_2 \psi^2 
+\delta_2 \psi^3 +\beta_2 \psi^4 +\gamma\phi^2 \psi^2+\eta \phi \psi^2\,,
\ee
where $\alpha_i, \delta_i, \beta_i$, $\gamma$ and $\eta_i$ are material (or 
system) dependent parameters.  Note that we have changed the notation slightly
from that followed in \cite{ks2}, in order to be in conformity with the
notation in the previous section. Hence the (static) equations of motion are  
\be\label{3.2}
\frac{d^2 \phi}{dx^2}=2\alpha_1 \phi +3\delta_1 \phi^2 +4\beta_1 \phi^3 
+2\gamma\phi \psi^2 +\eta \psi^2\,,
\ee
\be\label{3.3}
\frac{d^2 \psi}{dx^2}=2\alpha_2 \psi +3\delta_2 \psi^2 +4\beta_2 \psi^3 
+2\gamma\phi^{2} \psi +2\eta \psi \phi\,.
\ee
Observe that as long as $\eta \ne 0$, the two field equations are asymmetric
in $\phi$ and $\psi$. We shall consider solutions of these coupled field
equations in case $\alpha_i \ne 0, \delta_i \ne 0, \beta_i>0$, as only 
then the model corresponds to a first order transition in both the fields.

There is only one solution in this case.  It is not difficult to show that
\be\label{3.4}
\phi =F+A\sn^2[D(x+x_0),m]\,,~~\psi = G+B\sn^2[D(x+x_0),m]\,,
\ee
is an exact solution to coupled field Eqs. (\ref{3.2}) and (\ref{3.3}) provided 
the following eight field equations are satisfied
\be\label{3.5}
2\alpha_1 F+3\delta_1 F^2+4\beta_1 F^3+2\gamma FG^2+\eta G^2 = 2AD^2\,,
\ee
\be\label{3.6}
2\alpha_1 A+6\delta_1 AF+12\beta_1 F^2 A+4\gamma BFG+2\gamma AG^2
+2\eta BG = -4(1+m)AD^2\,,
\ee
\be\label{3.7}
3\delta_1 A^2+12\beta_1 F A^2+2\gamma FB^2+4\gamma ABG+\eta B^2 =6AmD^2\,,
\ee
\be\label{3.8}
2\beta_1 A^2+\gamma B^2=0\,,
\ee
\be\label{3.9} 
2\alpha_2 G+3\delta_2 G^2+4\beta_2 G^3+2\gamma GF^2+2\eta FG =2BD^2\,,
\ee
\be\label{3.10}
2\alpha_2 B+6\delta_2 BG+12\beta_2 G^2 B+4\gamma AFG+2\gamma BF^2
+2\eta(AG+BF) =-4(1+m)BD^2\,,
\ee
\be\label{3.11}
3\delta_2 B^2+12\beta_2 G B^2+2\gamma GA^2+4\gamma ABF
+2\eta AB =6mBD^2\,,
\ee
\be\label{3.12}
2\beta_2 B^2+\gamma A^2=0\,.
\ee
Five of these equations 
determine the five unknowns $A,B,D,F,G$, while the other three equations, 
give three constraints between the eight parameters $\alpha_{1,2},
\delta_{1,2},\beta_{1,2}, \gamma, \eta$. 
In particular, from the above equations it follows that 
\be\label{3.13}
\gamma<0\,,~~|\gamma|^2=4\beta_1 \beta_2\,, ~~~\sqrt{\beta_1}A^2
=\sqrt{\beta_2}B^2\,.
\ee

From Eq. (\ref{3.9}) it follows that no solution of the form (\ref{3.4}) 
exists in case $G=0$. Thus no solutions exist with $\psi=B\sn^2[D(x+x_0),m]$, 
irrespective of the value of $F$.
In fact one can also show that no solution exists even in case $B=-G$ or if
$B=-mG$ unless $m=1$. In other words, solutions of the form 
$\psi=G \cn^2[D(x+x_0),m]$ or $\psi=G\dn^2[D(x+x_0),m]$ do not exist, no 
matter what $F$ is except when $m=1$. 

{\bf Solution at $m=1$}: In the special case of $m=1$, the solution (\ref{3.4}) 
goes over to the hyperbolic nontopological soliton solution
\be\label{3.14}
\phi =F+A\tanh^2[D(x+x_0)]\,,~~\psi =G+B\tanh^2[D(x+x_0)]\,,
\ee
provided the field Eqs. (\ref{3.5}) to (\ref{3.12}) with $m=1$ are satisfied.
This hyperbolic soliton solution takes a particularly simple form in two cases
which we mention one by one. 

(i) {\bf $F=0,\, G=-B$}: In this limit the nontopological soliton 
solution (\ref{3.14}) takes the simpler form
\be\label{3.15}
\phi =A\tanh^2[D(x+x_0)]\,,~~\psi =B\sech^2[D(x+x_0)]\,.
\ee
By analyzing Eqs. (\ref{3.5}) to (\ref{3.12}) it is easily shown that such a
solution exists provided $\gamma <0\,,\alpha_2<0$. Further, while
Eqs. (\ref{3.8}) and (\ref{3.12}) still continue to hold good, the other field
equations take a slightly simpler form
\be\label{3.16}
2AD^2=\eta B^2\,,
\ee
\be\label{3.17}
6AD^2=3\delta_1 A^2+4|\gamma|AB^2+\eta B^2\,,
\ee
\be\label{3.18}
-4AD^2=\alpha_1 A-|\gamma| AB^2-\eta B^2\,,
\ee
\be\label{3.19}
2D^2=\alpha_2-|\gamma|A^2+\eta A\,,
\ee
\be\label{3.20}
6D^2=3\delta_2 B-4|\gamma|A^2+2\eta A\,.
\ee
From here one can easily solve for $A,B,D$ and further obtain four
constraints between the eight parameters.

(ii) {\bf $F=-A, G=-B$}: In this limit the nontopological soliton 
solution (\ref{1s}) takes the simpler form
\be\label{3.21}
\phi =A\sech^2[D(x+x_0)]\,,~~\psi =B\sech^2[D(x+x_0)]\, , 
\ee
provided field Eqs. (\ref{3.8}) and (\ref{3.12}) hold good
and further 
\be\label{3.22}
\alpha_1=\alpha_2>0\,,~~2D^2=\alpha_1\,,
\ee
\be\label{3.23}
-6AD^2=3\delta_1 A^2+\eta B^2\,,~~-6BD^2=3\delta_2 B^2+2\eta AB\,.
\ee

It turns out that as long as $\delta_2 \ne 0$, no other 
Lam\'e polynomials of order two 
form a solution of field Eqs. (\ref{3.2}) and (\ref{3.3}).

\section{Asymmetric-Symmetric $\phi^4$ Model: Lam\'e Polynomial 
Solutions of Order one}

In the last section we considered solutions in case both $\delta_1$ and
$\delta_2$ are nonzero, i.e. solutions of the asymmetric $\phi^4$ problem 
such that in both $\psi$ and $\phi$ fields one has a first order phase 
transition. In this section, we consider the case when $\delta_1 \ne 0$ 
while $\delta_2=0$. This corresponds to having a first order transition 
in $\phi$ and a second order transition in $\psi$. There are interesting 
physical situations such as a face-centered cubic to a hexagonal close 
packed (FCC-HCP) reconstructive structural transition \cite{rochal} and 
the martesnitic transition in cobalt \cite{toledano} where this model is 
relevant. Therefore, in this section we consider such a coupled model and 
obtain various solutions of this coupled model in terms of Lam\'e 
polynomials of order one, and their hyperbolic limit. In the next section, 
we shall show that the Lam\'e polynomials of order two are also the exact 
solutions of this coupled model, even though they are not the solutions 
in the decoupled limit.

The potential we consider is given by ($\beta_1>0,\beta_2 > 0$)
\be\label{4.1a}
V=\alpha_1 \phi^2 -\delta_1 \phi^3 +\beta_1\phi^4 +\eta\phi \psi^2
+\gamma\phi^2 \psi^2+\alpha_2 \psi^2+\beta_2\psi^4\,,
\ee
where $\alpha_{1,2},\beta_{1,2},\delta_1,\eta,\gamma$ are system dependent 
parameters. The static field equations that follow from here are
\be\label{4.2}
\phi_{xx}=2\alpha_1\phi-3\delta_1 \phi^2+4\beta_1\phi^3+\eta\psi^2
+2\gamma \phi \psi^2\,,
\ee
\be\label{4.3}
\psi_{xx}=2\alpha_2\psi+4\beta_2\psi^3+2\eta\phi \psi+2\gamma\phi^2 \psi\,.
\ee
Observe that as long as $\eta \ne 0$, the two field equations are asymmetric 
and hence kink-pulse and pulse-kink solitons would be distinct.

For the uncoupled model ($\eta=\gamma=0$), it is easy to show that the 
potential in $\phi$ corresponds to a first order transition while that in 
$\psi$ corresponds to a second order transition. In particular, while 
$\phi=0$ is the only minimum in case $\delta_1^2 < (32/9)\alpha_1\beta_1$, 
for $(32/9)\alpha_1\beta_1 < \delta_1^2 <4\alpha_1\beta_1$, $\phi=0$ is the 
absolute minimum while $\phi=\phi_c$ is the local minimum, whereas for 
$\delta_1^2>4\alpha_1\beta_1$, the opposite is true. At $\delta_1^2=4\alpha_1 
\beta_1$ we have degenerate minima at $\phi=0$ and $\phi=\phi_c$.  It is well 
known that while at $T=T^{I}_c$ (i.e. $\delta_1^2=4\alpha_1\beta_1$), one has 
a kink solution, for both  $T> T^{I}_c$ and $T<T^{I}_c$ one has a pulse 
solution around the local minimum.  On the other hand, while $\psi=0$ is the 
only extremum, i.e. minimum in case $\alpha_2 >0$ (i.e. $T>T_c^{II}$), for 
$\alpha_2<0$, $\psi=0$ is the maximum, while there are degenerate absolute 
minima at $\psi=\pm \psi_c$ (i.e. $T<T_c^{II}$). In this case, one has a kink 
solution for $T < T^{II}_{c}$. 

Let us now write down the periodic soliton solutions of this coupled 
asymmetric-symmetric model in terms of Lam\'e polynomials of order one. We 
shall see that  at $\delta_1^2=4\alpha_1\beta_1$ (corresponding to the 
uncoupled case), one has nine solutions in the coupled case. 

We look for the most general periodic solutions in terms of the Jacobi 
elliptic functions sn($x,m$), cn($x,m$) and dn($x,m)$ \cite{gr}.  
Since this model is almost similar to the asymmetric coupled $\phi^4$ 
model \cite{ks2}, except that while $\delta_2$ is nonzero in that case, 
in the present case
$\delta_2=0$, hence most of the results about the Lam\'e polynomial 
solutions are very similar in the two cases. We shall therefore only focus 
on those results which are different in the two cases.

{\bf Solution I}

It is not difficult to show that
\be\label{4.4}
\phi =F+A\sn[D(x+x_0),m]\,,~~\psi =G+B\sn[D(x+x_0),m]\,,
\ee
is an exact solution to coupled Eqs. (\ref{4.2}) and (\ref{4.3}) provided 
the following eight coupled equations are satisfied
\be\label{4.5}
2\alpha_1 F-3\delta_1 F^2+4\beta_1 F^3+\eta G^2+2\gamma FG^2=0\,,
\ee
\be\label{4.6}
2\alpha_1 A-6\delta_1AF+12\beta_1 F^2 A+2\eta BG+2\gamma AG^2
+4\gamma FBG =-(1+m)AD^2\,,
\ee
\be\label{4.7}
-3\delta_1A^2+12\beta_1 F A^2+\eta B^2+2\gamma FB^2+4\gamma ABG=0\,,
\ee
\be\label{4.8}
2\beta_1 A^2+\gamma B^2=mD^2\,,
\ee
\be\label{4.9} 
2\alpha_2 G+4\beta_2 G^3+2\eta FG+2\gamma GF^2 =0\,,
\ee
\be\label{4.10}
2\alpha_2 B+12\beta_2 G^2 B+2\eta AG+2\eta FB+4\gamma AFG+2\gamma BF^2
=-(1+m)BD^2\,,
\ee
\be\label{4.11}
12\beta_2 G B^2+2\eta AB+2\gamma GA^2+4\gamma ABF =0\,,
\ee
\be\label{4.12}
2\beta_2 B^2+\gamma A^2=mD^2\,.
\ee
Five of these equations 
determine the five unknowns $A,B,D,F,G$ while the other three equations, 
give three constraints between the seven parameters $\alpha_{1,2}, 
\beta_{1,2},\delta_1,\eta,\gamma$. In particular, $A$ and $B$ are given by
\be\label{4.13}
A^2=\frac{mD^2(2\beta_2-\gamma)}{(4\beta_1 \beta_2 -\gamma^2)}\,, ~~~ 
B^2=\frac{mD^2(2\beta_1-\gamma)}{(4\beta_1 \beta_2 -\gamma^2)}\,.
\ee

It is easily shown that, while no solution exists in case $F=0$, irrespective 
of whether $G$ is zero or nonzero, a solution exists  in case $G=0,F \ne 0$. 
In fact the analysis becomes somewhat simpler in that case and we shall 
restrict our discussion to that case.

\noindent{\bf G=0, F$\ne$0:}

In this case $A,B$ are again given by Eq. (\ref{4.13}) while $D$ and $F$ are 
given by 
\be\label{4.14}
D^2=\frac{\alpha_1}{(1+m)}\,,~~F=\sqrt{\frac{\alpha_1}{4\beta_1}}\,,
\ee
and further the following three constraints have to be satisfied
\be\label{4.15}
\delta_1^2=4\alpha_1\beta_1\,,~~~\eta^2=2\gamma(\alpha_1+2\alpha_2)\,,
~~~\delta_1 \eta+2\alpha_1 \gamma=0\,.
\ee
Thus this solution exists at $T=T_c^{I}$ for $\phi$ and $T<T_c^{II}$
for $\psi$.

\noindent{\bf m=1}

In this limiting case we have a kink-kink solution given by
\be\label{4.16}
\phi =F+A\tanh[D(x+x_0)]\,,~~\psi =B\tanh[D(x+x_0)]\,,
\ee
with $A$, $B$, $F$ and $D$ given by Eqs. (\ref{4.13}) and (\ref{4.14}) 
with $m=1$ while the three constraints are again given by Eq. (\ref{4.15}).  

{\bf Solution II}

A different type of solution (pulse lattice) is given by
\be\label{4.17}
\phi =F+A\cn[D(x+x_0),m]\,,~~\psi =G+B\cn[D(x+x_0),m]\,,
\ee
which is an exact solution provided eight coupled equations similar to
those given by Eqs. (\ref{4.5}) to (\ref{4.12}) are satisfied.
In particular, in this case $A$ and $B$ are given by
\be\label{4.18}
A^2=\frac{mD^2(2\beta_2+|\gamma|)}{(\gamma^2-4\beta_1 \beta_2)}\,, ~~~ 
B^2=\frac{mD^2(2\beta_1+|\gamma|)}{(\gamma^2-4\beta_1 \beta_2)}\,.
\ee
Notice that unlike the previous solution, this solution exists only when 
$\gamma<0$ and $\gamma^2>4\beta_1 \beta_2$. 

Unlike the previous case, it turns out that in this case a solution exists
both when $G=0,F \ne 0$ and $F=0, G \ne 0$ and we discuss both the cases one 
by one.

\noindent{\bf G=0, F$\ne$0:}

In this case $A,B$ are again given by Eq. (\ref{4.18}) while $D$ and $F$ are 
given by
\be\label{4.19}
D^2=\frac{\alpha_1}{(1-2m)}\,,~~F=\sqrt{\frac{\alpha_1}{4\beta_1}}\,.
\ee
Further, there are three constraints given by
\be\label{4.20}
\eta \delta_1=2|\gamma|\alpha_1\,,~~\delta_1^2=4\alpha_1\beta_1\,,
~~\eta^2+2|\gamma|\alpha_1=-4|\gamma|\alpha_2\,.
\ee 
Notice that this solution exists only if $\alpha_1>0,\alpha_2<0$ and $m<1/2$.
Thus this solution exists at $T=T_c^{I}$ for $\phi$ and $T<T_c^{II}$
for $\psi$.

\noindent{\bf F=0, G$\ne$0:}

In this case, solution (\ref{4.17}) exists only if $\eta=0$ and 
$\alpha_2<0,\gamma<0$.
While $A,B$ are again given by Eq. (\ref{4.18}), $D$ and $G$ are given by
\be\label{4.21}
D^2=\frac{4|\alpha_2|}{(2m-1)}\,,~~G=\sqrt{\frac{|\alpha_2|}{2\beta_2}}\,.
\ee
Further, there are three constraints given by
\be\label{4.22}
2\alpha_1\beta_2=|\alpha_2|(4\beta_2+|\gamma|)\,,~~3\delta_1 A+4|\gamma|BG=0\,,
~~|\gamma|A^2=6\beta_2 B^2\,.
\ee 
Notice that this solution exists only if $\alpha_1>0,\alpha_2<0$ and  $m>1/2$.
Thus this solution exists at $T=T_c^{I}$ for $\phi$ and $T<T_c^{II}$
for $\psi$.

\noindent{\bf m=1}

In this limiting case we have a pulse-pulse solution given by
\be\label{4.23}
\phi =A\sech[D(x+x_0)]\,,~~\psi =G+B\sech[D(x+x_0)]\,,
\ee
provided the relations (\ref{4.18}),  (\ref{4.21}) and (\ref{4.22}) are
satisfied with $m=1$.

{\bf Solution III}

Yet another pulse lattice solution is given by
\be\label{4.24}
\phi =F+A\dn[D(x+x_0),m]\,,~~\psi =G+B\dn[D(x+x_0),m]\,,
\ee
which is an exact solution provided eight coupled equations similar to 
Eqs. (\ref{4.5}) to (\ref{4.12}) are satisfied. 
In particular, $A$ and $B$ are given by
\be\label{4.25}
A^2=\frac{D^2(2\beta_2+|\gamma|)}{(\gamma^2-4\beta_1 \beta_2)}\,, ~~~ 
B^2=\frac{D^2(2\beta_1+|\gamma|)}{(\gamma^2-4\beta_1 \beta_2)}\,.
\ee
Thus this solution too exists only if $\gamma<0$ and 
$\gamma^2>4\beta_1\beta_2$. 

Unlike the solution II, this solution exists only if $F=0, G \ne 0$.  In this 
case, solution (\ref{4.24}) exists only if $\eta=0$ and $\alpha_2<0,\gamma<0$.
While $A,B$ are again given by Eq. (\ref{4.25}), $D$ and $G$ are given by
\be\label{4.26}
D^2=\frac{4|\alpha_2|}{(2-m)}\,,~~G=\sqrt{\frac{|\alpha_2|}{2\beta_2}}\,.
\ee
Further, there are three constraints given by Eq. (\ref{4.22}).

At $m=1$, this solution too goes over to the pulse-pulse solution
as given by Eq. (\ref{4.23}).

\noindent{\bf Solution IV}

We shall now discuss mixed solutions to the coupled Eqs. (\ref{4.2}) 
and (\ref{4.3}). It turns out that all these solutions exist only if
$G$ is necessarily zero while $F$ is necessarily nonzero. 

It is easily shown that
\be\label{4.27}
\phi =F+A\sn[D(x+x_0),m]\,,~~\psi =G+B\cn[D(x+x_0),m]\,,
\ee
is an exact solution provided 
\be\label{4.28}
G=0\,,~~ \eta+2\gamma F=0\,,~~\eta<0\,,~~2\beta_1>\gamma>2\beta_2>0\,.
\ee
In particular, $A$ and $B$ are given by
\be\label{4.29}
A^2=\frac{mD^2(\gamma-2\beta_2)}{(\gamma^2-4\beta_1 \beta_2)}\,, ~~~ 
B^2=\frac{mD^2(2\beta_1-\gamma)}{(\gamma^2-4\beta_1 \beta_2)}\,,
\ee
while $D$ is given by
\be\label{4.30}
(1+m)D^2=\alpha_1-2\gamma B^2\,.
\ee
Further, the three constraints are
\be\label{4.31}
\delta_1^2=4\alpha_1 \beta_1\,,~~2\alpha_1 \gamma+\eta \delta_1=0\,,
~~(2m-1)D^2=2\alpha_2+2\gamma (A^2-F^2)\,.
\ee

\noindent{\bf m=1}

In this limiting case we have a kink-pulse solution given by
\be\label{4.32}
\phi =F+A\tanh[D(x+x_0)]\,,~~\psi =B\sech[D(x+x_0)]\,,
\ee
with $A$, $B$, $F$ and $D$ given by Eqs. (\ref{4.28}) to (\ref{4.30}) 
with $m=1$
while the three constraints are again given by Eq. (\ref{4.31}).  

\noindent{\bf Solution V}

It is easy to show that another such solution is
\be\label{4.33}
\phi =F+A\sn[D(x+x_0),m]\,,~~\psi =G+B\dn[D(x+x_0),m]\,.
\ee
This is an exact solution provided Eq. (\ref{4.28}) is satisfied.
Further, $A$ and $B$ are given by
\be\label{4.34}
A^2=\frac{mD^2(\gamma-2\beta_2)}{(\gamma^2-4\beta_1 \beta_2)}\,, ~~~ 
B^2=\frac{D^2(2\beta_1-\gamma)}{(\gamma^2-4\beta_1 \beta_2)}\,,
\ee
while $D$ is again given by Eq. (\ref{4.30}).
In addition, two of the three constraints are again given by Eq. (\ref{4.31})
while the third one is now given by
\be\label{4.35}
(2-m)mD^2=2m\alpha_2+2\gamma(A^2-mF^2)\,.
\ee

In the limiting case of $m=1$, we again have the same kink-pulse solution 
as given by Eq. (\ref{4.32}).

\noindent{\bf Solution VI}

We now discuss two periodic solutions which at $m=1$ reduce to 
pulse-kink solutions. 
In particular, it is easily shown that
\be\label{4.36}
\phi =F+A\cn[D(x+x_0),m]\,,~~\psi =G+B\sn[D(x+x_0),m]\,,
\ee
is an exact solution provided Eq. (\ref{4.28}) is satisfied.
Further, $A$ and $B$ are given by
\be\label{4.37}
A^2=\frac{mD^2(\gamma-2\beta_2)}{(4\beta_1 \beta_2-\gamma^2)}\,, ~~~ 
B^2=\frac{mD^2(2\beta_1-\gamma)}{(4\beta_1 \beta_2-\gamma^2)}\,,
\ee
while $D$ is given by
\be\label{4.38}
(2m-1)D^2=2\gamma B^2-\alpha_1\,.
\ee
In addition, two of the three constraints are again given by Eq. (\ref{4.31})
while the third constraint is 
\be\label{4.39}
-(1+m)D^2=2\alpha_2+2\gamma (A^2-F^2)\,.
\ee

\noindent{\bf m=1}

In this limiting case we have a pulse-kink solution given by
\be\label{4.40}
\phi =F+A\sech[D(x+x_0)]\,,~~\psi =B\tanh[D(x+x_0)]\,,
\ee
with $A$, $B$, $F$ and $D$ given by Eqs. (\ref{4.37}) to (\ref{4.39}) 
with $m=1$.

\noindent{\bf Solution VII}

Another periodic solution which at $m=1$ reduces to the 
same pulse-kink solution (\ref{4.40}) is given by 
\be\label{4.41}
\phi =F+A\dn[D(x+x_0),m]\,,~~\psi =G+B\sn[D(x+x_0),m]\,,
\ee
provided Eq. (\ref{4.28}) is satisfied.
Further, $A$ and $B$ are given by
\be\label{4.42}
A^2=\frac{D^2(\gamma-2\beta_2)}{(4\beta_1 \beta_2-\gamma^2)}\,, ~~~ 
B^2=\frac{mD^2(2\beta_1-\gamma)}{(4\beta_1 \beta_2-\gamma^2)}\,,
\ee
while $D$ is given by
\be\label{4.43}
m(2-m)D^2=2\gamma B^2-2\alpha_1 m\,.
\ee
In addition, two of the three constraints are again given by Eq. (\ref{4.31})
while the third constraint is given by Eq. (\ref{4.39}).

{\bf Solution VIII}

We now discuss two mixed solutions which at $m=1$ reduce to pulse-pulse
solution. In particular, it is easy to check that
\be\label{4.44}
\phi =F+A\cn[D(x+x_0),m]\,,~~\psi =G+B\dn[D(x+x_0),m]\,,
\ee
is an exact solution to Eqs. (\ref{4.2}) and (\ref{4.3}) 
provided 
\be\label{4.45}
G=0\,,~~\eta+2\gamma F=0\,,~~\gamma<0\,,~~\gamma^2>4\beta_1 \beta_2\,.
\ee
Further, $A$ and $B$ are given by
\be\label{4.46}
A^2=\frac{mD^2(|\gamma|+2\beta_2)}{(\gamma^2-4\beta_1 \beta_2)}\,, ~~~ 
B^2=\frac{D^2(|\gamma|+2\beta_1)}{(\gamma^2-4\beta_1 \beta_2)}\,,
\ee
while $D$ is given by
\be\label{4.47}
(1-2m)D^2=2|\gamma|(1-m)B^2+\alpha_1\,.
\ee
In addition, two of the three constraints are again given by Eq. (\ref{4.31})
while the third constraint is given by 
\be\label{4.48}
(2-m)mD^2=2m\alpha_2+2|\gamma|[(1-m)A^2-mF^2]\,.
\ee
At $m=1$, the solution (\ref{4.44}) goes over to the pulse-pulse solution
\be\label{4.49}
\phi =F+A\sech[D(x+x_0)]\,,~~\psi =B\sech[D(x+x_0)]\,,
\ee
satisfying conditions (\ref{4.45}) to (\ref{4.47}) with $m=1$. 

{\bf Solution IX}

Another mixed solution is
\be\label{4.50}
\phi =F+A\dn[D(x+x_0),m]\,,~~\psi =G+B\cn[D(x+x_0),m]\,,
\ee
provided Eq. (\ref{4.45}) is satisfied.
Further, $A$ and $B$ are given by
\be\label{4.51}
A^2=\frac{D^2(|\gamma|+2\beta_2)}{(\gamma^2-4\beta_1 \beta_2)}\,, ~~~ 
B^2=\frac{mD^2(|\gamma|+2\beta_1)}{(\gamma^2-4\beta_1 \beta_2)}\,,
\ee
while $D$ is given by
\be\label{4.52}
m(2-m)D^2=2|\gamma|(1-m)B^2-m\alpha_1\,.
\ee
In addition, two of the three constraints are again given by Eq. (\ref{4.31})
while the third constraint is given by 
\be\label{4.53}
(2m-1)D^2=2\alpha_2-2|\gamma|[(1-m)A^2-F^2]\,.
\ee
At $m=1$, the solution (\ref{4.50}) also goes over to the pulse-pulse solution
(\ref{4.49}).

In addition to these, there are four other solutions which have been discussed 
in \cite{ks2} (i.e. in case $\delta_2 \ne 0$) which continue to be valid even 
when $\delta_2=0$. As an illustration, we discuss only one of these solutions 
at $m=1$ which in the uncoupled case corresponds to 
$\delta_1^2>4\alpha_1\beta_1$, i.e.  
$T<T_c^{I}$.

{\bf Solution X}

It is easily shown that
\be\label{4.54}
\phi =\frac{A\sech[D(x+x_0)]}{1+H\sech[D(x+x_0)]}\,,~~
\psi =\frac{B\sech[D(x+x_0)]}{1+H\sech[D(x+x_0)]}\,,
\ee
is an exact solution to the field Eqs. (\ref{4.2}) and (\ref{4.3}) 
provided $A$ and $H$ are given by
\be\label{4.55}
A^2=\frac{(H^2-1)D^2(2\beta_2-\gamma)}{(4\beta_1 \beta_2-\gamma^2)}\,, ~~~ 
B^2=\frac{(H^2-1)D^2(2\beta_1-\gamma)}{(4\beta_1 \beta_2-\gamma^2)}\,,  
\ee
and
\be\label{4.56}
D^2=2\alpha_1=2\alpha_2\,,~~\eta A+3\alpha_1 H=0\,,
~~\eta (\gamma+2\beta_1)+3\beta_1 \gamma=0\,.
\ee

\section{Asymmetric-Symmetric $\phi^4$ Model: Lam\'e Polynomial 
Solutions of Order Two}

In the last section we discussed the Lam\'e polynomial solutions of order 
one for the field Eqs. (\ref{4.2}) and (\ref{4.3}) which are also the
solutions of the uncoupled problem. We now show that the field 
Eqs. (\ref{4.2}) and (\ref{4.3}) also admit Lam\'e polynomial solutions
of order two, which are not the solutions of the uncoupled problem.

\subsection{Solution I}

It is not difficult to show that 
\be\label{5.1}
\phi =F+A\sn^2[D(x+x_0),m]\,,~~\psi = G+B\sn^2[D(x+x_0),m]\,,
\ee
is an exact solution to coupled field Eqs. (\ref{4.2}) and (\ref{4.3}) provided 
the field Eqs. (\ref{3.5}) to (\ref{3.12}) with $\delta_2=0$ are satisfied.
Five of these equations 
determine the five unknowns $A,B,D,F,G$, while the other three equations, 
give three constraints between the seven parameters $\alpha_{1,2},
\delta_{1},\beta_{1,2}, \gamma, \eta$. 
In particular, Eq. (\ref{3.13}) must be satisfied. Most of the conclusions 
drawn in the last section for this solution also apply in this case.

{\bf Solution at $m=1$}: In the special case of $m=1$, the solution (\ref{5.1}) 
goes over to the hyperbolic nontopological soliton solution (\ref{3.14}).
This hyperbolic soliton solution takes a particularly simple form in two cases
which we mention one by one. 

(i) {\bf $F=0,\, G=-B$}: In this limit the nontopological soliton 
solution (\ref{3.14}) takes the simpler form
\be\label{5.2}
\phi =A\tanh^2[D(x+x_0)]\,,~~\psi =B\sech^2[D(x+x_0)]\,.
\ee
By analyzing Eqs. (\ref{3.5}) to (\ref{3.12}) it is easily shown that such a
solution exists provided $\gamma <0\,,\alpha_2<0$. Further, while
Eqs. (\ref{3.8}) and (\ref{3.12}) still continue to hold good, the other field
equations take a slightly simpler form as given by Eqs. (\ref{3.16}) to
(\ref{3.19}). However, instead of Eq. (\ref{3.20}), we now have
\be\label{5.2a}
6D^2=-4|\gamma|A^2+2\eta A\,.
\ee

(ii) {\bf $F=-A, G=-B$}: In this limit the nontopological soliton 
solution (\ref{1s}) takes the simpler form
\be\label{5.3}
\phi =A\sech^2[D(x+x_0)]\,,~~\psi =B\sech^2[D(x+x_0)]\, , 
\ee
provided field Eqs. (\ref{3.8}) and (\ref{3.12}) hold good
and further 
\be\label{5.4}
\alpha_1=\alpha_2>0\,,~~2D^2=\alpha_1\,,
\ee
\be\label{5.5}
-6AD^2=3\delta_1 A^2+\eta B^2\,,~~-3D^2=2\eta A\,.
\ee

{\bf Solution II}

Unlike the previous section, it turns out that in view of $\delta_2 =0$,
now three more Lam\'e polynomial solutions of order two are allowed which we
present one by one.

It is not difficult to show that
\be\label{5.6}
\phi =F+A\sn^2[D(x+x_0),m]\,,~~\psi =B\sn[D(x+x_0),m]\cn[D(x+x_0),m]\,,
\ee
is an exact solution to coupled field Eqs. (\ref{4.2}) and (\ref{4.3}) provided 
the following seven field equations are satisfied
\be\label{5.7}
2\alpha_1 F+3\delta_1 F^2+4\beta_1 F^3 =2AD^2\,,
\ee
\be\label{5.8}
2\alpha_1 A+6\delta_1 AF+12\beta_1 F^2 A+2\gamma FB^2
+\eta B^2 =-4(1+m)AD^2\,,
\ee
\be\label{5.9}
3\delta_1 A^2+12\beta_1 F A^2+2\gamma B^2 (A-F)-\eta B^2
=6AmD^2\,,
\ee
\be\label{5.10}
2\beta_1 A^2-\gamma B^2=0\,,
\ee
\be\label{5.11} 
2\alpha_2 +2\gamma F^2+2\eta F =-(4+m)D^2\,,
\ee
\be\label{5.12}
4\beta_2 B^2+4\gamma AF
+2\eta A =6mD^2\,,
\ee
\be\label{5.13}
2\beta_2 B^2-\gamma A^2=0\,.
\ee
Four of these equations determine the four unknowns $A,B,D,F$ while the 
other three equations, give three constraints between the seven parameters 
$\alpha_{1,2},\beta_{1,2}, \gamma, \delta_1, \eta$. 
In particular, from Eqs. (\ref{5.10}) and (\ref{5.13}) it follows that 
\be\label{5.14}
\gamma>0\,,~~\gamma^2=4\beta_1 \beta_2\,, ~~~\sqrt{\beta_1}A^2
=\sqrt{\beta_2}B^2\,.
\ee
From Eq. (\ref{5.7}) it is clear that no solution exists in case $F=0$. In
fact, one can also show that no solution exists even if $A=-F$ or $A=-mF$
unless $m=1$.

{\bf Solution at $m=1$}: In the special case of $m=1$, the solution (\ref{5.6}) 
goes over to the hyperbolic nontopological soliton solution
\be\label{5.15}
\phi =F+A\tanh^2[D(x+x_0)]\,,~~\psi =B\tanh[D(x+x_0)]\sech[D(x+x_0)]\,,
\ee 
provided the field Eqs. (\ref{5.7}) to (\ref{5.13}) with $m=1$ are satisfied.

It is not difficult to show that unlike the solution (\ref{5.6}), the
solution
\be\label{5.16}
\phi =B\sn[D(x+x_0),m]\cn[D(x+x_0),m]\,,~~
\psi =F+A\sn^2[D(x+x_0),m]\,,
\ee
does not exist as long as $\delta_1$ and $\eta$ are nonzero. 

\subsection{Solution III}

Another allowed solution is
\be\label{5.17}
\phi =F+A\sn^2[D(x+x_0),m]\,,~~\psi =B\sn[D(x+x_0),m]\dn[D(x+x_0),m]\,,
\ee
provided seven field equations similar to Eqs. (\ref{5.7}) to (\ref{5.13})
are satisfied.
In particular, one can show that solution (\ref{5.17}) exists provided
\be\label{5.18}
\gamma>0\,,~~\gamma^2=4\beta_1 \beta_2\,, ~~~\sqrt{\beta_1}A^2
=m\sqrt{\beta_2}B^2\,.
\ee

{\bf Solution at $m=1$}: In the special case of $m=1$, the solution 
(\ref{5.17}) goes over to the hyperbolic nontopological soliton solution 
(\ref{5.15}).

\subsection{Solution IV}

Finally, another allowed solution is
\be\label{5.19}
\phi =F+A\sn^2[D(x+x_0),m]\,,~~\psi =B\cn[D(x+x_0),m]\dn[D(x+x_0),m]\,,
\ee
provided coupled equations similar to (\ref{5.7}) to (\ref{5.13}) are
satisfied.

In particular, one can show that such a solution exists only if
\be\label{5.20}
\gamma<0\,,~~|\gamma|^2=4m\beta_1 \beta_2\,, ~~~\sqrt{\beta_1}A^2
=\sqrt{m\beta_2}B^2\,.
\ee

{\bf Solution at $m=1$}: In the special case of $m=1$, the solution 
(\ref{5.20}) goes over to the hyperbolic nontopological soliton solution 
\be\label{5.21}
\phi =F+A\tanh^2[D(x+x_0)]\,,~~\psi =B\sech^2[D(x+x_0)]\,,
\ee 
which is essentially the solution (\ref{1s}). 

\section{Conclusion}

In this paper we have shown that the Lam\'e polynomials of order two are
the periodic solutions of the coupled $\phi^4$ problems when either the 
potentials in both the fields are symmetric, or when both are asymmetric 
or when the potential is symmetric in one and asymmetric in the other field. 
The latter model is also relevant for reconstructive phase transitions in 
many materials \cite{rochal,toledano}.  These are novel solutions in the 
sense that while they are the solutions of the coupled problems, they are 
not the solutions of the corresponding uncoupled problems. In particular, 
in all the three cases we have shown that while the Lam\'e polynomials of 
{\it order one} are the solutions of {\it both} the coupled and the 
uncoupled problems, the Lam\'e polynomials of {\it order two} are the 
solutions of the coupled problems, but {\it not} of the uncoupled ones.  

It is worth emphasizing here that there are three Lam\'e polynomials of 
order one and as a result one has nine different solutions for the coupled 
$\phi^4$ problems which we have presented in \cite{ks1,ks2} and in Sec. IV 
of this paper. Since there are five Lam\'e polynomials of order two one 
would have naively expected sixteen solutions of order two for the coupled 
$\phi^4$ problems [note that two of the Lam\'e polynomials are of the form 
$F+A\sn^2(x,m)$].  However, it turned out that while there are {\it six} 
allowed solutions in the symmetric $\phi^4$ case in an external field, only 
{\it one} solution is allowed in the asymmetric case and {\it four} solutions 
are allowed in the asymmetric-symmetric case.

It may be noted here that previously, Hioe and Salter \cite{hs} had shown 
similar features for coupled nonlinear Schr\"odinger (NLS) equations. They 
also pointed out that precisely when such solutions exist, the coupled NLS 
equations are integrable and they pass the Painlev\'e test \cite{rl}.  Thus 
one might get the impression that the existence of higher order Lam\'e 
polynomials as solutions of the coupled problem (but not that of the 
uncoupled problem) could be related to the integrability of the  coupled as 
well as the uncoupled systems. However, our work has clearly shown that this 
is not so.  In particular, it is well known that the $\phi^4$ field theory 
(both symmetric or asymmetric or mixed) is a nonintegrable field theory.

As a further support to our argument, we consider elsewhere \cite{ks5} 
a coupled $\phi^6$ model studied by us recently \cite{ks4}, and show that 
provided we add extra interaction terms which are quadratic-quartic in the 
two fields, then Lam\'e polynomials of order two are also the solutions of 
the coupled problem (though they are not the solutions of the uncoupled 
problem).

Based on these examples, we conjecture that for most of the coupled models,
novel solutions (i.e. those which are solutions of the coupled but not the 
uncoupled problem) will exist as long as there is a coupling term between
the fields which is of the same order as the highest power of the uncoupled
fields. Further, in those cases where Lam\'e polynomials, of say order one, 
are solutions of the uncoupled problem, we conjecture that if there are 
$n$-coupled fields with coupling terms being of the same order as the highest 
power of the uncoupled fields, then Lam\'e polynomials of order $n$ will also  
be the solutions of the coupled problem.  Note that four coupled $\phi^4$ 
fields are required to model different magnetic phases of the hexagonal 
multiferroic materials \cite{munawar}.  It will be interesting to examine 
our conjecture in a few coupled field theory models. We hope to address these 
issues in future. 

\section{Acknowledgment}
This work was supported in part by the U.S. Department of Energy.

\end{document}